\documentclass{tlp}

\usepackage{harvard}
\usepackage{epsfig}

\usepackage{latexsym}

\newtheorem{example}{Example}[section]

\newcommand{\uea}{{update and access}}
\newcommand{\ii}{{\hspace*{1cm}}}
\newcommand{\iii}{{\hspace*{2cm}}}
\newcommand{\iiii}{{\hspace*{3cm}}}
\newcommand{\iiiii}{{\hspace*{4cm}}}
\newcommand{\ia}{{\hspace*{0.8cm}}}

\begin{document}

\bibliographystyle{dcu}

\title{An Open Ended Tree}
\author[Henk Vandecasteele, Gerda Janssens]{
\begin{tabular}{cc}
HENK VANDECASTEELE & GERDA JANSSENS\\
  PharmaDM                        & Katholieke Universiteit Leuven, \\
                                  & Department of Computer Science\\
 Ambachtenlaan 54D                & Celestijnenlaan 200A \\
 B-3001 Leuven, Belgium           & B-3001 Leuven, Belgium \\
{Henk.Vandecasteele@PharmaDM.com} &
{ Gerda.Janssens@cs.kuleuven.ac.be}
\end{tabular}
}

\newcommand\othrpearl{P\ls R\ls O\ls G\ls R\ls A\ls M\ls M\ls I\ls  N\ls G\ns P\ls E\ls A\ls R\ls L}

\maketitle[o]

\shorttitle{Programming pearl}

\begin{abstract}
An open ended list is 
a well known
data structure in Prolog programs.
It is frequently used to represent a value changing over time, while this value
is referred to from several places in the data structure of the
application. 
A weak point in this technique is that
the time complexity 
is linear
in the number of updates to the value represented by the open ended list. 
In this 
{\em programming pearl} we present a variant of the open ended
list, namely an open ended tree, with an {\uea} time complexity 
logarithmic in the number of updates to the value. 
\end{abstract}

\section{Introduction}
Many applications in Logic Programming deal with 
variables of which the content changes
over time.
In this programming pearl these variables are called
the {\em application variables}.
An example of such an application is a  Constraint Logic Programming 
Finite Domain solver (CLP(FD)) \cite{Chip88}.
In such a solver
the {\em application variables} are the finite domain variables.
The solver changes the domains of the finite domain variables and also
the set of constraints associated with the finite domain variables.
Another example is found in a fix-point process that computes
subsequent approximations for an entity before a final value -- the
fix-point -- is reached. 
It is often the case that several entities depend on each other.
The application variables are the entities
for which a fix-point has to be computed.
The {\em application variables}
are updated and used in an interleaved way.  The number of 
{\em application variables} is not known in advance.

The problem is to find a representation 
for such {\em application variables} that
are updated and accessed in an interleaved and unpredictable way.
Several solutions exist to tackle this problem: 

\begin{itemize}
\item{\bf Replacement}\\
In this solution every occurrence of the application 
variable in the data structure must be replaced on every change of
the content of the
{\em application variable}. 
For simple applications this might be feasible, but in a complex 
application as a CLP Finite Domain solver this is not 
reasonable, since each finite domain variable can occur in
numerous constraints
and the solver has to traverse
all these constraints at each change in the domain of
a finite domain variable.

\item{\bf Threading}\\
Threading a pair of arguments, containing the current values 
of the {\em application variables}, is usually seen
as the most 
obvious solution. 
The first argument then contains the incoming
state, the current values at the time of the call. The
second argument contains the set of values as the resulting state of the call.
All occurrences of the {\em application variables}  
in other data structures simply 
refer to the values in these states (e.g. by some numbering scheme).
Although preferable from logical point of view, it can be problematic
from efficiency point of view. When dealing with a large number of
{\em application variables}, 
the access and update is at least logarithmic in
the number of {\em application variables}  
(e.g. when stored in a balanced tree). 
In the case of a demanding application with many values to be 
maintained, such as a Finite Domain solver, 
this extra time complexity is
significant.

If the {\em application variables} are known beforehand, the programmer
can thread a corresponding number of pairs through the program and
avoid the search for the value.  In our examples, the number of
{\em application variables} is different for each use of the program.

\item{\bf Open ended lists}\\
The use of open ended lists
avoids dependency on the number of {\em application variables}
in the application
and also avoids replacing each occurrence of them whenever the value
changes.
The rule is
that the last element before the open end is the current value of the
{\em application variable}. 
Whenever the same {\em application variable} occurs 
in a data structure of
the application, the same open ended list is referred to. 
Whenever the value of the {\em application variable}  changes,
the end of the list is instantiated to a new list with the 
new value as first element and with a new open end.
In this way all the other occurrences of the same {\em application variable}
can see the change.
When storing an already existing {\em application variable} one can
use a list that consists
only of the last element and the open end.
(e.g. [a,b,c,d$|$Var] can be replaced by [d$|$Var]).
The target applications in this pearl do not often allow this replacement.
 
As every application variable is represented by a separate open ended
list,
update and access times 
do not depend on the number of application
variables.  Every update to an application variable adds one element
to the open ended list: the length of the list is equal to the number
of updates to the application variable.  From the observations that
in order to add a value one has to instantiate the tail and
that the current value is the last element in the open ended list, we
can conclude that the \uea\ time is linear in the  number of
updates to the {\em application variable}.
From data complexity point of view, open ended lists have a serious
disadvantage compared to threading: the technique
will keep  {\em alive} all
values that are in the list. This means that the garbage collector
will never be able to collect any of the values used in the past.

\item{\bf Assert and retract}\\
Assert and retract
can be used to store the changing content over time.
This  method has a time complexity for \uea\ which is 
independent of the number 
of {\em application variables} and the number of updates.  
One may have to deal with a high constant factor in 
most Prolog systems. 
Every lookup requires the creation of a new instance of the value.
When the values of the {\em application variables} 
contain logic variables, this method  will not work
because  every lookup will return an instance with fresh variables. 
When
large values are used, it may lead to a lot of overhead due to the
creation and the garbage collection of the instances. 
In all other solutions
discussed here, no new instances are created when accessing the
current value.
When dynamic predicates are used, old values will not be restored on
backtracking  as it is the case for  open ended lists or threading. 
Assigning a new value is now destructive
and old values can be removed
if the Prolog system has a garbage collector for dynamic code.

\item{\bf Non portable solutions}\\
Some Prolog systems provide their own solution based on
  backtrackable destructive assignment \cite{plilp92,plilp90},
for example as attributed variables \cite{SICSTUS} and meta variables \cite{eclipse98}.
Using these features 
is probably efficient but unfortunately not portable.
The \uea\ times for these solutions are O(1).
Also data complexity is optimal in these solutions: old values that
are not kept alive by some choice point can be collected by the
heap garbage collector. 
\end{itemize}

In this programming pearl we present the {\bf open ended tree} as an
alternative data structure for representing {\em application variables} that
are updated and accessed in an interleaved and unpredictable way.
The open ended tree is an ISO-compatible solution and has an 
\uea\
complexity which is logarithmic in the
number of updates 
to the {\it application variable} at hand.
The data complexity is equivalent to the data complexity of the 
open ended list solution.

In Section 2 we explain how an open ended tree is used to represent an
{\em application variable}.  Section 3 gives the Prolog predicates for
accessing and updating the value of an {\em application variable}.
Section 4 shows some efficiency results and 
presents some variants that can be used to tune the application at hand.

\section{Open ended trees}

In an open ended tree, the current value of the
{\em application variable} is found 
in the rightmost leaf of the tree: 
it is the last instantiated node 
that would be encountered when the tree were traversed in a depth-first left-to-right way.
The tree is constructed such that
the number of steps for finding this rightmost leaf is logarithmic 
in the number of nodes in the tree. This number of nodes is
the same as the number of updates.

The main issue is the shape of the tree. Since the
number of updates is not known in advance and the nodes of the tree 
can not be rearranged, a balanced tree is out of the question.
The solution is to create 
a sequence of
binary trees, where each tree is one 
level deeper than the previous tree. This 
sequence of trees could
be stored in an open ended list, but for simplicity of the lookup-procedure
the binary tree structure is reused.

The nodes that are used to build the 
sequence of trees are called the
{\bf collector nodes}. The right child of a collector node is --
if already created -- again a collector node.
A new collector node can only be created if the left child of the 
parent
has reached depth N in all its branches. The left child
of the newly created collector node is restricted to depth N+1.
Each node in the open ended tree contains
two children and a data field. 
A child
can be a free variable or again a node. Such a free
variable can be instantiated later with a node, as is done in open ended lists.
The root node is a collector
node, whose left child's depth is limited to 1.  An empty open ended
tree is represented by a free variable. 

\begin{figure}
\begin{center}
\epsfig{file=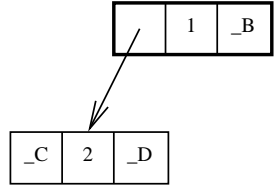,  height=2.5cm}
\end{center}
\caption{An open ended tree with 1 collector node and a tree of depth 1}
\label{fig1}
\end{figure}

\begin{figure}
\begin{center}
\epsfig{file=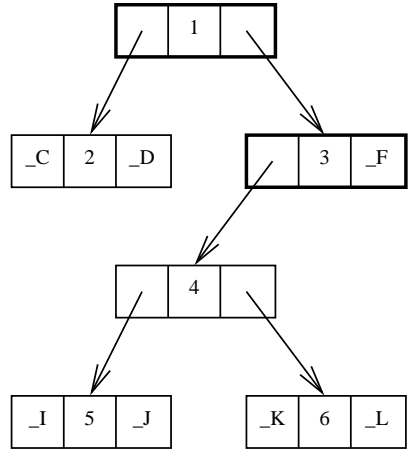,  height=6cm}
\end{center}
\caption{An open ended tree with  2 collector nodes and a tree of depth 1 and depth 2}
\label{fig2}
\end{figure}

\begin{example} 
Suppose the open ended tree O is used to represent an 
{\em application   variable} whose subsequent values are 1, 2, 3, $\ldots$, 10.
These values will be added one after another to O.
Firstly,  ``1'' is added and O gets bound to 
\vspace{-0.1cm}
\begin{quote}
{\em {\bf tree(}\_A,{\bf 1},\_B{\bf )}}, 
\end{quote}
\vspace{-0.1cm}
a {\bf collector node}  with free variables as children.
\label{one-value}
The left child becomes a binary tree of depth 1 when ``2'' is added:
\vspace{-0.1cm}
\begin{quote}
{\em {\bf tree(}tree(\_C,2,\_D),{\bf 1},\_B{\bf )}}.
\end{quote}
\vspace{-0.1cm}
This open ended tree is shown in Figure \ref{fig1}. 
Note that
collector nodes are put in bold in the text.
When adding the third value ``3'', a new {\bf collector node} is created as
the right child of the root collector node:
\vspace{-0.15cm}
\begin{quote}
{\em{\bf tree(}tree(\_C,2,\_D),{\bf 1},{\bf tree(}\_E,{\bf 3},\_F{\bf ))}}.
\end{quote}
\vspace{-0.15cm}
Adding ``4'', a tree of depth 2 is started:
\vspace{-0.15cm}
\begin{quote}
{\em {\bf tree(}tree(\_C,2,\_D),{\bf 1},
{\bf tree(}tree(\_G,4,\_H),{\bf 3},\_F){\bf ))}}.
\end{quote}
\vspace{-0.15cm}
After  ``5'' and ``6'' have been added, the tree of depth 2 is 
completed as shown in Figure \ref{fig2}:
\vspace{-0.1cm}
\begin{quote}
{\em 
{\bf tree(}tree(\_C,2,\_D), {\bf 1},
{\bf tree(}tree(tree(\_I,5,\_J),4,tree(\_K,6,\_L)),{\bf 3},\_F{\bf ))}
}.
\end{quote}
\vspace{-0.1cm}
Adding all values up to ``10'' gives the following open ended tree:
\vspace{-0.1cm}
\begin{quote}
{\em {\bf tree(}tree(\_C,2,\_D), {\bf 1},
{\bf tree(}tree(tree(\_I,5,\_J),4,tree(\_K,6,\_L)), {\bf 3},\\
{\bf tree(}tree(tree(tree(\_T,10,\_U),9,\_S),8,\_Q),{\bf 7},\_O{\bf )))}
}.
\end{quote}
\vspace{-0.1cm}
\end{example}

The use of a set of trees to reduce the access time to some data
structure is not new,  e.g. 
Fibonacci heaps~\cite{fredman}. 
However the shape and the properties of open ended trees
are quite different: open ended trees are binary and have a different shape,  and no
reorganisation of the trees is ever needed to assure 
logarithmic time complexity 
for the operations we need.

We can prove
that the \uea\ to the data structure are
logarithmic in the number of updates.
A tree of depth N contains maximum
$\sum^{N-1}_{i=0}{2^i} = 2^N -1$ nodes. 
Since the collector node contains a value as well, we have 
$2^N$ values in a tree of depth $N$ and its corresponding 
collector node.
After the tree of depth N has been completed, the
data structure contains $\sum_{i=1}^{N}2^i = 2^{N+1} -2$ elements.
After the tree of depth N-1 is completed and the tree of depth N is 
under construction the data structure contains M nodes where
\begin{equation}
(2^{N} - 2) < M \leq (2^{N+1} -2) 
\end{equation}
M is also the number of updates.
From (1) we deduce that
$N = ceil(log_2(M+2)) -1$.  
Then finding the last tree takes N steps and 
finding the rightmost leaf in this tree takes at most N steps as
well. This results in a time complexity of $O(log(M))$.

The main disadvantage of this approach compared to an open 
ended list is its space consumption: it uses twice as much memory
in most Prolog implementations. 
An open ended list uses one ./2 term for each element in the list.
A ./2 term needs two heap cells, whereas 
an open ended tree has per element one tree/3 term which takes 4 heap
cells.

\section{Code for the open ended tree}
In this section we give the Prolog predicates that define the
two operations on an open ended tree that represents an 
{\em application variable}:
\begin{itemize}
\item
lookup(ApplVar, Value) unifies Value with the current value of the application
variable represented by ApplVar.
\item
insert(ApplVar, New) stores New as the (updated) current value of the
application variable represented by ApplVar.
\end{itemize}
Subsequent calls of insert(T, X) and lookup(T, Y) will always unify the
two variables X and Y.\\
\% lookup(T,V) finds the current value V in the rightmost leaf of the tree T\\
lookup(tree(Left, El, Right), Value):-\\
\ii   (  var(Right) $\rightarrow$\\ 
\iii      (  var(Left) $\rightarrow$ El = Value 
      ;
         lookup(Left, Value) 
      )\\
\ii  ;
\ia      lookup(Right, Value)\\
\ii   ).\\
\% insert(T,V) stores the current value V in a new node in the tree T as the\\
\% rightmost leaf \\
insert(Tree, Value):-\\
\ii   (  nonvar(Tree) $\rightarrow$
         insert1(Tree, Value, 1) \\
\ii   ;
\ia      Tree = tree(\_, Value, \_) \% the first collector node\\
\ii   ).\\
\% insert1(T,V,D) first finds the last collector node in T and meanwhile computes\\
\% the depth D of the tree in the left branch of T. Next it inserts the value V\\
\% in the left branch of the collector node. In case this tree is full, insert1 creates\\
\% a new collector node in the right child of the node in variable T. \\
insert1(tree(Left, \_, Right), Value, Depth):-\\
\ii   (  var(Right) $\rightarrow$ \\
\iii      insert2(Left, Value, Depth, Right)\\
\ii   ;
\ia      Depthplus1 is Depth + 1,\\
\iii      insert1(Right, Value, Depthplus1)\\
\ii   ).\\
\% insert2(T,V,D,R) inserts V in the tree T, unless this would make the depth of T\\
\% larger than D. In the latter case a node is created in the variable R. This variable\\ 
\% R is known to be the next subtree to be instantiated, it could be a collector node.\\
insert2(tree(Left, El, Right), Value, Depth, Back):-\\
\ii   (  var(El) $\rightarrow$ El = Value\\
\ii   ;
\ia      (  Depth == 1 $\rightarrow$ Back = tree(\_, Value, \_)\\
\iii      ;
\ia         Depthmin1 is Depth - 1,\\
\iiii         (  var(Right) $\rightarrow$\\
\iiiii            insert2(Left, Value, Depthmin1, Right)\\
\iiii         ;
\ia            insert2(Right, Value, Depthmin1, Back)\\
\iiii         )\\
\iii      )\\
\ii   ).
\section{Efficiency results and optimisations}

\subsection{Measuring efficiency}

Our benchmarks measure the difference in efficiency between open ended
lists and open ended trees, which are both portable solutions that can be
used in the same kind of circumstances.

Four experiments were done, each starting from a data structure with 
already 
several updates. In the first experiment we started with a 
data structure with already 10 updates; the subsequent experiments had
a data structure with already 100, 1000 and 100000 updates.
Then, in each of these experiments, the time to update the data structure,
and the time to access the current value was measured.
Each of the operations (update and access) was repeated 100000 times
(In case of update, the update was undone by backtracking
to prevent the data structure from growing).
Each of the experiments was done on an implementation with open ended
lists and open ended trees. 
The computation was performed on a Pentium III 666 Mhz, running Linux 2.2.20,
both with ilProlog(version 0.9.6)~\cite{cw295} and
SICStus(version 3.9.0)~\cite{SICSTUS}. 
The times are reported in seconds and do not include the time for
setting
up the benchmark.
\begin{table}
\begin{center}
\begin{tabular}{|r|r|r|r|r|r|r|r|r|r|}
\hline
\# updates & \multicolumn{4}{c|}{ilProlog} & \multicolumn{4} {c|}{SICStus} & Depth\\

     & \multicolumn{2}{c|}{updating} &
       \multicolumn{2}{c|}{lookup} &
       \multicolumn{2}{c|}{updating} &
       \multicolumn{2}{c|}{lookup} & of tree\\

     & \multicolumn{1}{c|}{list} & \multicolumn{1}{c|}{tree} 
     & \multicolumn{1}{c|}{list} & \multicolumn{1}{c|}{tree} 
     & \multicolumn{1}{c|}{list} & \multicolumn{1}{c|}{tree} 
     & \multicolumn{1}{c|}{list} & \multicolumn{1}{c|}{tree} & \\
\hline
  10 & 0.07 & 0.16 & 0.07 & 0.08 & 0.18 & 0.32 & 0.19 & 0.22 & 6 \\
 100 & 0.53 & 0.28 & 0.51 & 0.15 & 1.50 & 0.62 & 1.82 & 0.41 & 12 \\
1000 & 4.88 & 0.39 & 4.85 & 0.20 &14.67 & 0.90 &18.11 & 0.53 & 18 \\
10000&49.86 & 0.52 &48.82 & 0.27 &147.1 & 1.29 &181.2 & 0.82 & 26 \\
\hline
\end{tabular}
\end{center}
\caption{\label{tab:time}Comparison of the execution times.}
\end{table}


From the experiments we can see that 
with only 10 updates the overhead is larger than the benefit of using open ended trees:
the disadvantage is rather small for lookup,
but considerable for update.
From 100 updates on, the overhead 
is compensated.
Furthermore, the timings for the open ended tree exhibit the expected 
logarithmic behaviour. The timings for the open ended list show
linear behaviour.
  
\subsection{Variants}
\begin{itemize}
\item
When using open ended lists, one can always replace the list by
some tail of the list,  as long as the tail contains at least one element
(e.g [1,2,3,4$|$V] can be replaced by [4$|$V]). 
Although this does not  change the time complexity of the resulting 
program, still a considerable speed-up can be realised.
The same technique can be used with open ended trees, after a 
modification of the code above. This modification consists of
storing the depth of the tree at each collector node. When this information
is available at the collector node, the root node can always be replaced
by one of the lower collector nodes. 
Optionally one can choose to put the depth in each node, such that 
computing depth while inserting can be avoided.

\item
When memory consumption is an issue, all leaves of the 
tree can be replaced by a smaller term\footnote{Thanks to a referee
  mentioning this}, e.g. leaf(value), or even simply 
the value if it is known to be 
nonvar.
The leaves of an open ended tree are all the nodes that occur at the
maximal depth in the trees.
In ilProlog a tree/3 term takes 4 heap cells, whereas a leaf/1
term
takes only 2. 
As on average half of the values are leaves and we gain 2 cells per
  leaf, the gain will be 1 heap cell per value.  If the value is
  stored directly in the leaf, 2 heap cells per value can be gained,
  and we have almost the same memory consumption as with open ended lists.
These observations are confirmed by our experiments. 

\item
When known in advance that {\em application variables} will have 
many updates,
one can start with larger trees in the 
sequence (e.g depth 10).
This speeds up the updates/lookups as can be seen in Table \ref{tab:timedepth}.
The timings are obtained with ilProlog.
\begin{table}
\begin{center}
\begin{tabular}{|r|r|r||r|r|}
\hline
\#updates &\multicolumn{2}{c||}{ Starting depth = 1} &
           \multicolumn{2}{c|}{ Starting depth = 10}\\ 
 & updating & lookup & updating & lookup \\
\hline
10     & 0.16  & 0.08 & 0.27  & 0.11\\
100    & 0.28 & 0.15 & 0.32 & 0.15\\
1000   & 0.39 & 0.20 & 0.31 & 0.12\\
10000  & 0.52 & 0.27  & 0.42 & 0.20\\
100000 & 0.66 & 0.34 & 0.57 & 0.29\\
\hline
\end{tabular}
\end{center}
\caption{\label{tab:timedepth} Starting the sequence of trees with a larger depth}
\end{table}


\end{itemize}

\end{document}